\documentstyle[epsfig]{mn}
\begin{document}
\LARGE
\normalsize

\title[GRS~1915+105]
{Infrared synchrotron oscillations in GRS~1915+105}
\author[R.~P.~Fender \& G.~G.~Pooley]
{R. P. Fender$^1$\thanks{email : rpf@astro.uva.nl}\thanks{EC Marie
Curie Fellow}
and G. G. Pooley$^2$\thanks{email : ggp1@cam.ac.uk}\\
$^1$ Astronomical Institute `Anton Pannekoek', University of Amsterdam,
and Center for High Energy Astrophysics, Kruislaan 403, \\
1098 SJ, Amsterdam, The Netherlands\\
$^2$ Mullard Radio Astronomy Observatory, Cavendish Laboratory,
Madingley Road, Cambridge CB3 OHE\\
}

\maketitle

\begin{abstract}

We report simultaneous observations of the black hole candidate X-ray
transient GRS~1915+105 in the infrared at K and in the radio at 2~cm.
Oscillations of period 26~min were observed in both wavebands,
having (dereddened) peak--peak amplitudes of about 40~mJy and with the
IR leading the radio by 7~min, or perhaps by 33 or 59~min.
A synchrotron origin for the oscillations continues to seem very likely.
We consider a range of problems raised by these observations, and
briefly discuss the applicability of expanding-synchrotron
source or conical jet models to the oscillations.
Comparing simplistic estimates of the ejecta mass to
the missing inner disc mass in the model of Belloni et al., we find
that a significant fraction of the inner disc may be ejected during
the oscillations.

\end{abstract}


\begin{figure*}
\centering
\leavevmode\epsfig{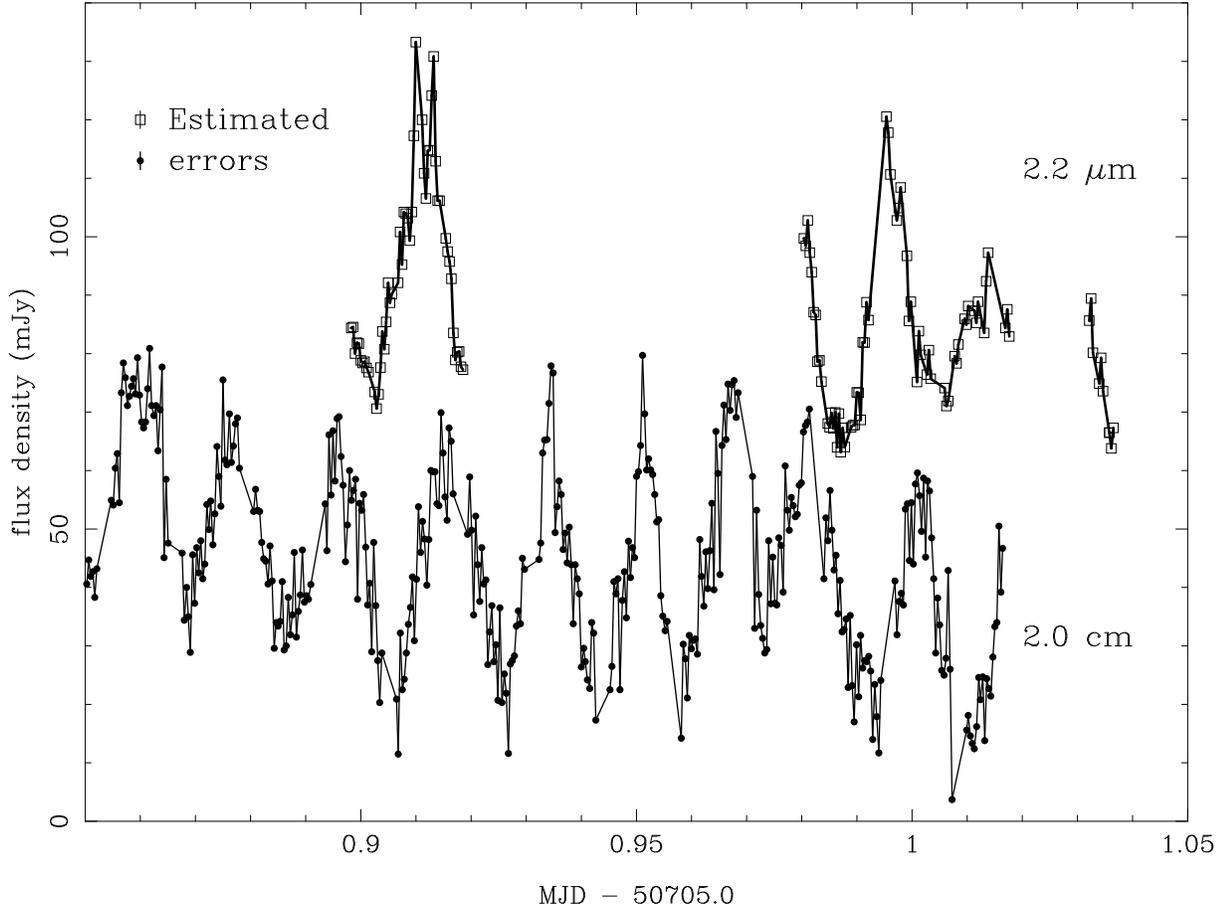}
\caption{Simultaneous radio (2~cm) and infrared K-band
(2.2~$\mu$m) light curves of GRS~1915+105. The infrared data
have been dereddened by $A_{\rm K} = 3.3$~mag. The oscillations
are clearly very similar across 4 decades of energy, and seem
to originate in a common population of synchrotron-emitting
electrons. The emission at 2.2~$\mu$m appears to lead that at 2~cm
by about 7 or 33~min, depending on which radio peak is identified
with the IR peak. The estimated errors for each data point
are indicated.}
\label{}
\end{figure*}

\begin{figure*}
\leavevmode{\epsfig{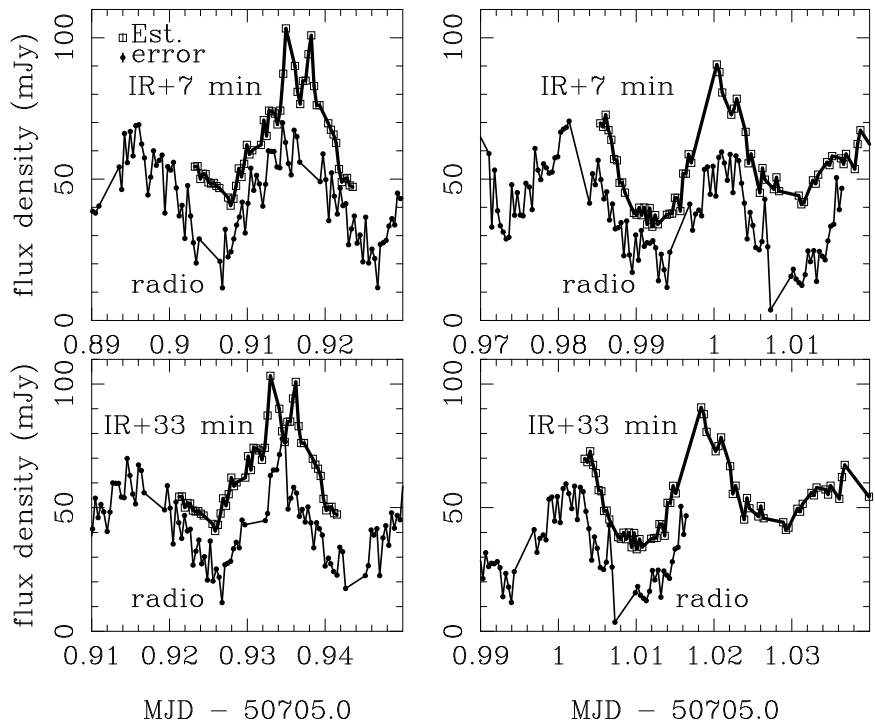}}
\caption{Details of two infrared oscillations and the nearest
radio events. The infrared data have been delayed by
7~min (upper plots) and 33~min (lower), and a constant 30~mJy
subtracted in order to facilitate comparison.
The data do not make a clear distinction between the two possible
delays, but they are not consistent with the opposite sense of delay
(radio variations ahead of the infrared) if the delay is constant.
The estimated errors for each frequency are indicated.}
\end{figure*}

\section{Introduction}

GRS~1915+105 is a highly unusual, extremely bright and variable black
hole candidate. Discovered as an X-ray source in 1992 (Castro-Tirado
et al. 1992) it has remained more or less detectable (although with an
extreme range of variability) for the past five years.  Shortly after
its discovery, Mirabel \& Rodr\'\i guez (1994) made observations of
the radio counterpart with the VLA and discovered jet-like outflows
with apparent superluminal motions corresponding to true bulk
velocities of $\sim 0.9c$.  As well as long-term variability, the
source exhibits a wide range of phenomena on time-scales from
milliseconds upwards, which may hold the key to the nature of the
accretion flow around the black hole. In particular, multiwavelength
oscillations with with periods 1--60 minutes have been the subject of
much discussion, as summarised below.

\subsection{Oscillations from GRS~1915+105}

Pooley (1995, 1996) reported the discovery of quasi-periodic radio
oscillations, with periods in the range 25 -- 40 min, from GRS
1915+105 in radio observations at 15 GHz.  Pooley \& Fender (1997),
hereafter PF97, reported over 18 months' of radio monitoring,
including many epochs of such radio QPO with periods in the range 20 -
60 min. On one occasion noted in that paper it was observed that
oscillations at 2~cm lead those at 3.6~cm by 4 - 5 min.  Rodr\'\i guez
\& Mirabel (1997) have also reported a sinusoidal radio oscillation.
Fender et al. (1997) discovered infrared K-band oscillations with a
similar amplitude and period and interpreted these as high-frequency
synchrotron counterparts to the radio oscillations, originating in
repeated small ejections of plasmons from the system (the first
suggestion of infrared synchrotron emission associated with the source
was made by Sams, Eckart \& Sunyaev (1996) from imaging observations
of an infrared jet).

Quasi-periodic dips and flickering were also observed in X-rays,
primarily with the XTE PCA (e.g. Greiner, Morgan \& Remillard 1996;
Morgan, Remillard \& Greiner 1997).  These have been interpreted by
Belloni et al.  (1997a, 1997b) as due to repeated disappearance and
refilling of the inner few hundred km of the accretion disc. Radio
observations simultaneous with PCA observations in 1996 October (PF97)
revealed that the radio oscillations are related in phase to the major
X-ray cycles, and hence with the disappearance of the inner disc
according to the Belloni model.  From these multiwavelength
observations, the idea was formed that during repeated disappearances
of the inner accretion disc, the majority of which may be advected
into the black hole, some fraction of the inner disc was accelerated
and ejected from the system and observed as synchrotron-emitting
plasmons (Fender et al. 1997; PF97).

Recently this general outline appears to have been confirmed by
simultaneous infrared/X-ray (Eikenberry et al. 1998, hereafter E98), and
infrared/X-ray/radio (Mirabel et al. 1998, hereafter M98) observations
of oscillations from GRS 1915+105. In particular,  E98, in high
time-resolution infrared observations, have revealed a
strong relation between infrared and X-ray flares, though at times the
emission in the two bands can decouple.  M98 report a radio and
infrared oscillation in which wavelength-dependent delays appear to be
compatible with a van der Laan (1966) model for synchrotron emission
from an ejected plasmon, and which occurs shortly after an X-ray dip.

\section{Observations}

GRS~1915+105 was observed between 20:34 and 01:19 on 1997 September
14/15 UT with the WHIRCAM infrared camera on the William Herschel
Telescope (WHT) on La Palma. Observations were made primarily through
the Ks (`short K') filter, with an effective wavelength of
2.2~$\mu$m, as well as a small number with the H filter (1.6~$\mu$m).
Standard G21-15 was used for calibration at K \& H, for which we
assumed magnitudes of 11.76 \& 11.85 respectively.  Conditions were
not photometric, but accurate photometry was performed relative to
star `A' as in Fender et al. (1997).

Radio observations at 15~GHz were made with the Ryle Telescope
(Cambridge) as part of our ongoing monitoring of the system. The
observing procedure is described more fully in PF97. Inspection of
flux densities at 2 and 8~GHz from the Green Bank Interferometer
monitoring program confirm
the presence of a variable source with an approximately flat radio spectrum
from 2 -- 15~GHz at the epoch of our observations.

We again deredden the infrared data by $A_{\rm K} = 3.3$~mag, but
caution that the H-band data, combined with J- and H-band observations
some 24 hr earlier, suggest that this may be a significant
overestimate of the extinction to the source (these results will be
discussed elsewhere). The simultaneous K-band infrared (dereddened)
and 2-cm radio observations are plotted in Fig 1. Estimated errors
on each data point are similar for the radio and dereddened infrared
data, at around 3~mJy r.m.s., and are indicated on the figure.
The relatively long gap in the IR observations resulted from an
instrumental problem while changing wavebands.

\section{Discussion}

We have observed IR and radio emission patterns which have strikingly similar
shapes and flux densities. It is also known that events of this sort are
closely related to the X-ray emission (PF97, M98, E98). There are many
questions raised about these phenomena, including the following:

(a) Is the IR radiation synchrotron emission?

(b) How are the shapes of the pulses in IR and radio defined?

(c) Is the near-equality of the flux density at 2.2~$\mu$m and 15 GHz a
coincidence?

(d) What are the phase relationships between the emission in the various
wavebands, and how do these relate to the underlying causes of the emission?

(e) Why does the X-ray emission appear to decouple from that in other
wavebands?

(f) Does the IR and/or the radio emission occur in material which is being
ejected from the system, and if so are the geometries and velocities similar
to those of the `superluminal' ejections first reported by Mirabel
\& Rodr\'\i guez (1994)?

We consider some of these issues briefly here.

\subsection {The IR emission mechanism}

The similarity between the IR and radio emission profiles and flux densities
suggests that the mechanism, assumed to be synchrotron emission for the radio,
is the same (Fender et al 1997). E98 argue that, because of the way in which
the IR and X-ray emissions decouple, the IR  cannot represent reprocessed
X-ray emission. A convincing IR polarimetric observation during these
quasi-periodic oscillations, and/or the detection of non-thermal brightness
temperatures (perhaps via very rapid variations?) would add support to the
synchrotron model.

\subsection {Pulse shape and phase relationships}

The pulse shape in the radio regime appears most frequently to have a short
($<$ 5 min) risetime and a somewhat slower decay, typically 10 -- 20 min (PF97)
with some variation between events. The IR data from E98, which have a higher
time-resolution, generally appear to conform to this picture, although there
are also some faster changes in the flux density.

The time-scales may be influenced by any or all of the following phenomena:
\begin{enumerate}
\item{the rate of supply of energy to the emission region;}
\item{light travel time across the emission region;}
\item{radiative losses of the emitting particles;}
\item{expansion of the emission region, with consequent changes in
optical depth, particle energy and magnetic field strength.}
\end{enumerate}

M98 apply the last of these in the form of a `van der Laan' (1966) expanding
synchrotron emission region. In this class of model, an expanding volume
of synchrotron-emitting material becomes optically thin at progressively
lower frequencies while the electrons and magnetic field both lose energy.
>From one event (1997 May 15) for which
they have infrared and three-frequency radio data M98 use the observed
ratios of the peak radio flux densities to derive an index {\it p}
for the distribution of the electron energies ($N(E) \propto E^{-p}$)
and delay from the initial injection of energy.
The delay is consistent with the timing of the preceding IR peak,
and the index derived is near zero,
which is substantially harder than values often found
for synchrotron sources, and would suggest a much greater IR flux density
than that observed. It is also difficult to reconcile the similarity of the IR
and 15-GHz pulse rise-times, which on this model are determined by the changes
in opacity, with the expectation that the synchrotron region would be optically
thin to IR radiation from the time of the initial injection.

Other authors have invoked jet-like geometries to account for the spectra and
variations of variable sources: Hjellming \& Johnston (1988) formulated a
conical-sheath model for the context of X-ray binaries, and others have
considered similar problems in relation to the variable, flat-spectrum emission
from the cores of AGN (e.g. Marscher \& Gear 1985, O'Dell 1988). Variants of
such models may apply here, but until we have some better idea of the geometry
of the outbursts in GRS~1915+105, application of these models may be
speculative. It appears that some expanding-source model will be required to
fit the radio-frequency data, but it is not yet clear what combination of
physics defines the shape of the IR pulses and why it is so similar to that of
the radio.

More simultaneous observations, including all of X-ray, IR and radio, are
needed to understand how the different emissions are related.

It is important to establish the delay, or range of delays, between
the infrared and radio peaks. M98 report a delay (2.2~$\mu$m to
3.6~cm) of 40 min on 97 May 15; and 16 min on 1997 Sep 09. We note,
however, that there are few cycles of data for any of the infrared
measurements, and the measurement of the delay is ambiguous; that on
1997 Sep 09, when the radio oscillations had a period of 40 min, might
be 56 min. Similarly, our observations on 1997 Sep 14-15 (radio period
26 min) are compatible with a delay of 7~min (Fig 2), or possibly 33
or 59 min (2.2~$\mu$m to 2~cm). The delay of 33 min gives a
marginally better subjective fit, while with a delay of 59~min there is
no longer any overlap for the second group of IR data. We assume
always that the infrared precedes the radio emission; an attempt
to fit our data with the delay reversed,
that is 15-GHz pulses preceding the IR,
is not successful unless different delays apply for the two
sections of IR data.
Identification of the range of observed delays
can only be done with extended observations when the source is in a
less regular state (for example, when a change from one period to
another is seen, or isolated pulses occur -- see examples in PF97 for
the radio, and the infrared data for 1997 Aug 14-15 in E98).

It also appears from the limited data so far that some of these delays vary
between observations. Does this reflect differing properties of the various
emission regions? Or is it perhaps a geometrical effect, where the apparent
delay changes with the relative orientation of some emitting structure and the
line of sight?  Determination of the true infrared -- radio delay, 
and its range of variation, will be crucial in modelling the oscillations.

\subsection {Ejection?}

Observations to determine the proper motions of these transient features do
not yet appear feasible. A search for spectral lines in the IR would be
difficult, and may show nothing if the emission is dominated by the
synchrotron process. A naive application of ``minimum energy'' arguments
to a synchrotron source, though subject to many caveats (e.g. Leahy 1991),
with a size defined by the 10-min
light-travel time and assuming one proton for every electron, yields a rest-mass
of the order of $10^{20} $~g -- comparable with the total mass
(about $10^{21}$~g)
estimated to be removed from the disc in one 25-min cycle in the model of
Belloni et al (1997b).

\section{Conclusions}

We have presented simultaneous infrared and radio observations of GRS
1915+105 demonstrating the similarity of oscillations, in period,
shape and amplitude, over more than 4 decades of energy. The radio
observations appear to lag the infrared by 7~min plus some multiple
of the oscillation period. Coupled with the
observations of Fender et al. (1997) and M98, there seems little doubt
that we are observed synchrotron oscillations between (at least) 15~GHz
and 2.2~$\mu$m from GRS 1915+105.

However, a clear model of the emission from radio to infrared
wavelengths, and its relation to the X-ray oscillations, has still not
been established.
Expanding synchrotron source models may
reproduce the observed emission and delays, depending on what we
accept as the true delay between emission at infrared and radio
wavelengths. Simultaneous observations at intermediate
(i.e. submillimetre) wavelengths may help to clarify the situation.
Simple van der Laan models as applied by M98 do not immediately
explain the similarity of flux densties and shapes of the pulses
in the IR and radio regimes.

\section*{Acknowledgements}

We thank Shaun Hughes for taking time out of his own PATT program to
implement our override observations, Tomaso Belloni and Michiel van
der Klis for stimulating discussions, and Ben Stappers for help with
IRAF.  We thank the staff at MRAO for maintenance and operation of the
Ryle Telescope, which is supported by the PPARC. The WHT is operated
on the island of La Palma by the Isaac Newton Group in the Spanish
Observatorio del Roque de los Muchachos of the Instituto de
Astrofisica de Canarias. RPF was supported during the period of this
research initially by ASTRON grant 781-76-017 and subsequently by EC
Marie Curie Fellowship ERBFMBICT 972436.

\end{document}